\documentclass[conference, a4paper]{IEEEtran}
\IEEEoverridecommandlockouts
\usepackage{adjustbox}

	\usepackage[turkish]{babel}
	\usepackage[utf8]{inputenc} 
	\usepackage[T1]{fontenc}

\usepackage{cite}

\usepackage{graphicx}

\usepackage[cmex10]{amsmath}
\usepackage{multirow}
\usepackage{array}
\usepackage[lofdepth,lotdepth]{subfig}
\usepackage{wrapfig}
\AtBeginDocument{%
  \renewcommand\tablename{TABLO}
}

\AtBeginDocument{%
  \renewcommand\abstractname{Abstract}
}

\begin{document}

\title{Verimli Bir Polip Bölütleme Ağı\\
An Efficient Polyp Segmentation Network}

\author{{Tuğberk Erol}\\
{Bilgisayar Mühendisliği} \\
{Gazi Üniversitesi}\\
{Ankara, Türkiye} \\
{tugberk.erol@gazi.edu.tr}
\and
{Duygu Sarıkaya} \\
{Bilgisayar Mühendisliği} \\
{Gazi Üniversitesi}\\
{Ankara, Türkiye} \\
{duygusarikaya@gazi.edu.tr}}

\maketitle

\begin{ozet}
Kanser hücrelerin kontrolsüz bir şekilde bölünüp çoğalması sonucu ortaya çıkan bir hastalıktır. Dünyada en sık görülen kanser türlerinden biri kalın bağırsak kanseridir. Kalın bağırsaktaki polipler erken müdahale ile alınmazsa kansere sebep olabilmektedir. Uzmanlar tanı sırasında gözden kaçan polipleri minimize etmek için derin öğrenme ile görüntü bölütleme tekniklerini kullanmaktadır. Bu teknikler iyi sonuçlar vermesine karşın çok fazla parametre gerektirmektedir. Bu soruna çözüm için yeni bir model öneriyoruz. Önerdiğimiz model, literatürde kabul görmüş çalışmalardan daha başarılı sonuçlar elde etmesinin yanı sıra daha az parametre gerektirmektedir. Modelimizde parametre sayısını azaltırken bölütlemede başarıyı korumak için kısmi kod çözücü kullanılmıştır. Modelimiz az parametre gerektirmesinin yanında başarılı sonuçlar elde eden EfficientNetB0 kodlayıcıya sahiptir. Polipler değişken şekil ve büyüklüğe sahip olduğu için klasik evrişim bloğu kullanmak yerine asimetrik evrişim bloğu kullanılmıştır. Bölütleme başarısını artırmak için özellik haritalarını ağırlıklandırarak önemli özellikleri ön plana çıkartan sıkma ve uyarma bloğu kullanılmıştır. Modelimizin başarısını ölçmek için Kvasir ve CVC-ClinicDB veri kümeleri eğitim, doğrulama ve test için kullanılırken, CVC-ColonDB, ETIS ve Endoscene veri kümeleri test için kullanılmıştır. Modelimiz Dice metriğine göre ColonDB test veri kümesinde \%71.8, EndoScene veri kümesinde \%89.3, ETIS veri kümesinde \%74.8 ile en iyi sonuçları almıştır. Modelimiz toplamda 2.626.337 parametre gerektirmektedir. Literatürde kıyasladığımız benzer çalışmalardan en az parametre gerektiren model 9.042.177 parametre sayısı ile U-Net++’dır.

\end{ozet}
\begin{IEEEanahtar}
Bilgisayarla Görü, Tıbbi Görüntü Bölütleme, Evrişimsel Sinir Ağları, Polip Bölütleme, Derin Öğrenme
\end{IEEEanahtar}

\begin{abstract}
Cancer is a disease that occurs as a result of the uncontrolled division and proliferation of cells. Colon cancer is one of the most common types of cancer in the world. Polyps that can be seen in the large intestine can cause cancer if not removed with early intervention. Deep learning and image segmentation techniques are used to minimize the number of polyps that goes unnoticed by the experts during these interventions. Although these techniques perform well in terms of accuracy, they require too many parameters. We propose a new model to address this problem. Our proposed model requires fewer parameters as well as outperforms the state-of-the-art models. We use EfficientNetB0 for the encoder part, as it performs well in various tasks while requiring fewer parameters. We use partial decoder, which is used to reduce the number of parameters while achieving high accuracy in segmentation. Since polyps have variable appearances and sizes, we use an asymmetric convolution block instead of a classic convolution block. Then, we weight each feature map using a squeeze and excitation block to improve our segmentation results. We used different splits of Kvasir and CVC-ClinicDB datasets for training, validation, and testing, while we use CVC-ColonDB, ETIS, and Endoscene datasets for testing. Our model outperforms state-of-art models with a Dice metric of \%71.8 on the ColonDB test dataset, \%89.3 on the EndoScene test dataset, and \%74.8 on the ETIS test dataset while requiring fewer parameters. Our model requires 2.626.337 parameters in total while the closest model in the state-of-the-art is U-Net++ with 9.042.177 parameters.

\end{abstract}

\IEEEpeerreviewmaketitle

\IEEEpubidadjcol

\section{G{\footnotesize İ}r{\footnotesize İ}ş}

Medikal görüntülerde bölütleme, tıbbi görüntülerde uzmanlar tarafından belirlenen bölgelerin diğer bölgelerden ayrılması işlemidir. Medikal görüntülerde bölütleme, uzmanlık gerektiren bir iş olduğundan maliyetli ve zaman alıcı bir işlemdir. Hatalı teşhis ve bölütleme hayati risk taşımaktadır. Kolorektal kanser, kadınlarda en sık görülen ikinci, erkeklerde en sık görülen üçüncü kanser türüdür\cite{kvasir}. Kolorektal kanser teşhisi, kalın bağırsağın incelenmesi ve buradaki poliplerin tespiti ile yapılır. Çalışmalar kolonoskopi sırasında poliplerin büyüklüğüne bağlı olarak gözden kaçma oranının \%14-30 arasında olduğunu göstermektedir\cite{kvasir}. Bunun için tanı sırasında poliplerin gözden kaçma oranını azaltmak, hastanın iyileşmesi için son derece önemlidir. Derin öğrenmedeki ilerlemeler, tıp alanında da kendisini göstermiştir. Derin öğrenmenin tıbbi görüntü bölütlemede kullanılması uzmanlara tanı koymada yardımcı olmakta ve teşhis başarısını arttırmaktadır.  Buna karşın derin öğrenme modelleri yüksek sayıda parametre gerektirmektedir. Bu durum kolonoskopi, endoskopi gibi gerçek zamanlı uygulamalarda sorun oluşturmaktadır. Bu sorunu çözmek için az parametre gerektiren yeni bir model öneriyoruz. Önerdiğimiz model,  kısmi kod çözücü kullanmasının yanı sıra, farklı boyutlardaki poliplerin başarılı bir şekilde bölütlenmesini sağlayan asimetrik evrişim bloğu ve özellik haritalarını ağırlıklandırarak önemli özellikleri ön plana çıkartan sıkma ve uyarma bloğu kullanmaktadır. Modelimiz literatürde kabul görmüş çalışmalardan daha başarılı sonuçlar elde etmesinin yanı sıra daha az parametre gerektirmektedir. Bildirinin II. bölümünde ilgili çalışmalara, bu çalışmaların dezavantajlarına ve modelimizde bu dezavantajları nasıl giderdiğimize, III. bölümde önerilen model ve kullanılan veri kümelerinin detaylarına, IV. bölümde ise deneyler, bu deneylerin sonuçlarına ve değerlendirmelere yer verilmiştir.   

\begin{figure*}[!h]
\centering
{\shorthandoff=%
\includegraphics[scale = 0.63]{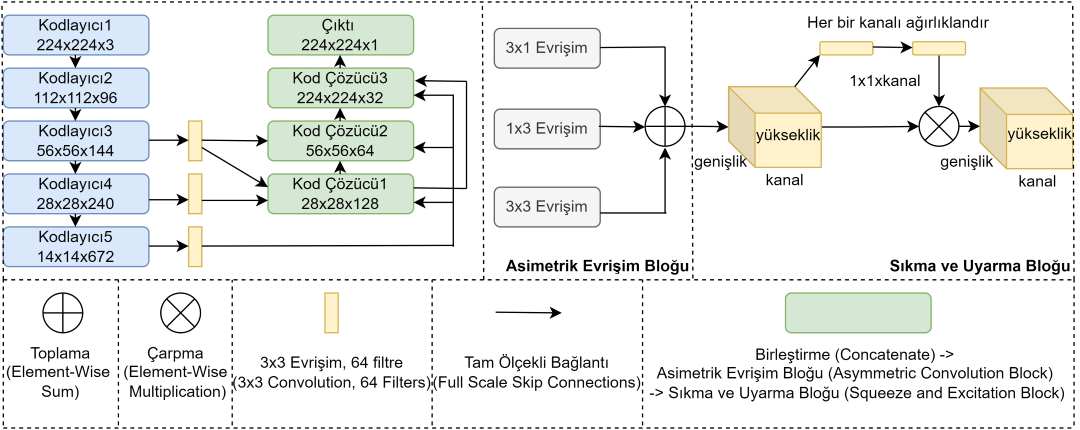}}
\caption{Çalışmamızda geliştirdiğimiz model. İlk bölümde parameter sayısını azaltmak için kullanılan kısmi kod çözücü yapısı ve farklı ölçeklerdeki bilgileri yakalamak için kullanılan tam ölçekli bağlantı görülmektedir. Altta her bir işaretin ne anlama geldiği gösterilmiştir. Sağ kısımda ise  asimetrik blok ve her bir kanalı ağırlıklandırıp önemli özellik haritalarını ön plana çıkartan sıkma ve uyarma blok yapısı gösterilmiştir.}
\label{model}
\end{figure*}

\section{{\footnotesize İ}lg{\footnotesize İ}l{\footnotesize İ} çalışmalar}
      Medikal görüntülerde bölütleme çalışmaları 2015 yılından sonra ivme kazanmıştır. 2015 yılında Ronneberger vd.\cite{unet} U-Net isimli bir biyomedikal görüntü bölütleme modeli önermişlerdir. Bu model günümüzde geliştirilen çoğu görüntü bölütleme modellerine ışık tutmaktadır. Jha vd.\cite{resunet++} ResUNet++ isimli bir medikal görüntü bölütleme modeli önermişlerdir. Kaybolan eğim problemine çözüm için Zhang vd. artık öğrenmeye dayalı\cite{resnet} ve U-Net\cite{unet} benzeri kodlayıcı ve kod çözücüye sahip bir yapı önermişlerdir. Modeli Kvasir-SEG\cite{kvasir} ve CVC-ClinicDB\cite{clinicdb}  veri kümelerinde test etmişlerdir. Kvasir-SEG 1000 görüntü, CVC-ClinicDB 612 görüntü içermektedir. Her iki veri kümesindeki görüntüler 256x256 boyutuna sahip olacak şekilde yeniden boyutlandırılmıştır. Her iki veri kümesi de \%80 eğitim, \%10 doğrulama ve \%10 test olacak şekilde ayrılmıştır. Modelin aşırı öğrenmesini engellemek için eğitim için ayrılan görüntülerde veri çoğaltma işlemleri yapılmıştır. Eğitim sonucuna göre bu model Kvasir-SEG test görüntülerinde \%81.33 Dice, \%79.27 IoU başarısı göstermiştir. CVC-ClinicDB’de ise \%79.55 Dice, \%79.62 IoU başarısı göstermiştir. Jha vd.\cite{dunet} Double U-Net isimli bir model  önermişlerdir. Bu modeldeki temel motivasyon, görüntülerden daha etkili bilgiler çıkarabilmek için iki U-Net ağının birbirine bağlanmasıdır. Kodlayıcı kısımda ön eğitimli VGG-19\cite{vgg} kullanılmıştır. İlk ağın çıktısı bir sonraki ağın girdisi olacak şekilde iki U-Net ağını birbirlerine bağlamışlardır. Çalışmayı CVC-ClinicDB veri kümesinde test etmişlerdir. Veri kümesi \%80 eğitim, \%10 doğrulama, \%10 test olarak ayrılmıştır. Görüntüler 384x512 boyutuna sahip olacak şekilde yeniden 
boyutlandırılmıştır. Önerilen model \%92.39 Dice, \%86.11 IoU başarısı göstermiştir. Fan vd.\cite{pranet} PraNet isimli bir model önermişlerdir. Bu çalışma paralel kısmi kod çözücü kullanmaktadır. Bu sayede model başarısına katkısı az ama parametre sayısını artıran yüksek çözünürlüklü kodlayıcı katmanlar alınmamış ve parametre sayısı azaltılmıştır. Poliplerin yapısal detaylarını daha iyi yakalamak için tersine dikkat mekanizması kullanmışlardır. Çalışma beş farklı veri kümesinde denenmiştir. Bu veri kümeleri Kvasir [1], CVC-ClinicDB\cite{clinicdb}, CVC-ColonDB\cite{colondb}, EndoScene\cite{endoscene} ve ETIS\cite{etis}’dir. Kvasir ve CVC-ClinicDB \%80 eğitim, \%10 doğrulama, ve \%10 test olarak ayrılmıştır. Kalan üç veri kümesi test olarak alınmıştır. Sonuçlar U-Net\cite{unet}, U-Net++\cite{unet++}, ResUNet++\cite{resunet++} gibi çalışmalarla kıyaslanmıştır. PraNet tüm metriklerde daha üstün sonuçlar elde etmiştir. Bu çalışmalarda önerilen modeller iyi sonuçlar vermesine rağmen karşın, yüksek sayıda parametre gerektirmektedir. Bu durum gerçek zamanlı tanı ve bölütleme çalışmalarında bir dezavantaj oluşturmaktadır. Çalışmamızda az parametre içeren ancak aynı zamanda yüksek doğrulukta başarı gösteren EfficientNetB0\cite{efficientnet} kodlayıcı kısımda kullanılmıştır. Wu vd.\cite{pd} ilk kodlayıcı katmanların model başarısına katkısının daha az olmasının yanında hesaplama maliyeti olarak da yüksek olduğunu belirtmişlerdir. Bunun yanında üçüncü katmandaki kodlayıcı katmanların birinci ve ikinci katmanın içerdiği düşük seviye bilgiye sahip olduğunu göstermişlerdir. Bu bilgi doğrultusunda, modelimizde parametre sayısını artıran ancak bölütleme başarısına etkisi az olan ilk iki kodlayıcı katmanı tam ölçekli bağlantıda kullanmamayı seçerek parametre sayısını önemli ölçüde azalttık. Polipler farklı şekil ve boyutlarda olduğundan klasik evrişim bloğu kullanmak yerine asimetrik evrişim bloğu kullandık. Dahası, özellik haritalarını ağırlıklandırarak önemli özellikleri ön plana çıkartan sıkma ve uyarma\cite{squeeze} bloğunu kullandık. Şekil \ref{model}’de modelimizin genel yapısı görülmektedir.

\begin{table}[!b]
\centering
\caption{METOTLAR VE PARAMETRE SAYILARI}
\begin{adjustbox}{scale=1.25,center}
\begin{tabular}{|c|c|}
\hline
\textbf{Metot}   & \textbf{Parametre Sayısı} \\ \hline
U-Net {[}2{]}    & 15.683.713                \\ \hline
U-Net++ {[}12{]} & 9.042.177                 \\ \hline
SFA {[}18{]}     & -                         \\ \hline
PraNet {[}8{]}   & 30.328.272                \\ \hline
\textbf{Bizimki} & \textbf{2.626.337}        \\ \hline
\end{tabular}
\end{adjustbox}
\label{parametre_sayısı}
\end{table}

\begin{table*}[t]
\caption{Geliştirdiğimiz model ve literatürdeki modeller IoU ve Dice metriğine göre kıyaslanmış ve tabloya eklenmiştir. Her bir metriğe göre en iyi sonucu veren modeller koyu renkte gösterilmiştir.}
\begin{adjustbox}{scale=1.5,center}
\begin{tabular}{|l|ll|ll|ll|ll|ll|}
\hline
\multirow{2}{*}{Metotlar} & \multicolumn{2}{l|}{\textbf{Kvasir}}                 & \multicolumn{2}{l|}{\textbf{ClinicDB}}               & \multicolumn{2}{l|}{\textbf{ColonDB}}                & \multicolumn{2}{l|}{\textbf{EndoScene}}              & \multicolumn{2}{l|}{\textbf{ETIS}}                   \\ \cline{2-11} 
                          & \multicolumn{1}{l|}{Dice}           & IoU            & \multicolumn{1}{l|}{Dice}           & IoU            & \multicolumn{1}{l|}{Dice}           & IoU            & \multicolumn{1}{l|}{Dice}           & IoU            & \multicolumn{1}{l|}{Dice}           & IoU            \\ \hline
UNET  {[}2{]}             & \multicolumn{1}{l|}{0.818}          & 0.746          & \multicolumn{1}{l|}{0.823}          & 0.755          & \multicolumn{1}{l|}{0.512}          & 0.444          & \multicolumn{1}{l|}{0.710}          & 0.627          & \multicolumn{1}{l|}{0.398}          & 0.335          \\ \hline
UNET++{[}12{]}            & \multicolumn{1}{l|}{0.821}          & 0.743          & \multicolumn{1}{l|}{0.794}          & 0.729          & \multicolumn{1}{l|}{0.483}          & 0.410          & \multicolumn{1}{l|}{0.707}          & 0.624          & \multicolumn{1}{l|}{0.401}          & 0.344          \\ \hline
SFA {[}18{]}              & \multicolumn{1}{l|}{0.723}          & 0.611          & \multicolumn{1}{l|}{0.700}          & 0.607          & \multicolumn{1}{l|}{0.469}          & 0.347          & \multicolumn{1}{l|}{0.467}          & 0.329          & \multicolumn{1}{l|}{0.297}          & 0.217          \\ \hline
PRANET {[}8{]}            & \multicolumn{1}{l|}{\textbf{0.898}} & \textbf{0.840} & \multicolumn{1}{l|}{0.899}          & 0.849          & \multicolumn{1}{l|}{0.709}          & \textbf{0.640} & \multicolumn{1}{l|}{0.871}          & 0.797          & \multicolumn{1}{l|}{0.628}          & 0.567          \\ \hline
\textbf{BİZİMKİ}          & \multicolumn{1}{l|}{0.894}          & 0.808          & \multicolumn{1}{l|}{\textbf{0.923}} & \textbf{0.857} & \multicolumn{1}{l|}{\textbf{0.718}} & 0.560          & \multicolumn{1}{l|}{\textbf{0.893}} & \textbf{0.806} & \multicolumn{1}{l|}{\textbf{0.748}} & \textbf{0.597} \\ \hline
\end{tabular}
\end{adjustbox}
\label{sonuclar}
\end{table*}

\begin{figure}[!t]
\centering
{\shorthandoff=%
\includegraphics[scale = 0.9]{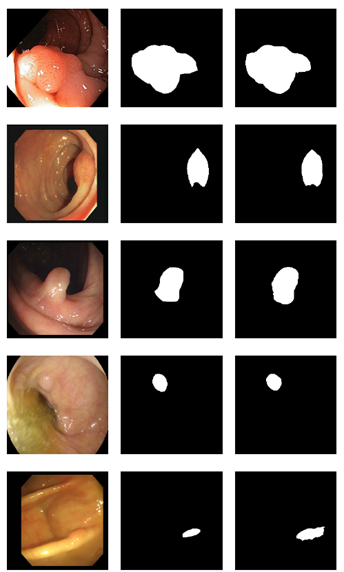}}
\caption{Test kümesinde yapılan tahmin işleminin sonucu. Her sıra soldan sağa doğru girdi görüntüsü, orijinal maske ve tahmin edilen maskeyi göstermektedir.Görüntüler yukarıdan aşağıya Kvasir, CVC-ClinicDB, ColonDB, Etis ve Endoscene veri kümelerinden alınmıştır.}
\label{polyps}
\end{figure}

\section{yöntem}
\subsection{Veri Kümesi ve Ön İşleme}
Çalışmamızda kullandığımız veri kümelerinden Kvasir\cite{kvasir}, CVC-ClinicDB\cite{clinicdb} \%80 eğitim, \%10 doğrulama ve \%10 test olarak ayrılmıştır. Eğitim ve doğrulamada 1450 görüntü, test için ise 100 Kvasir ve 61 CVC-ClinicDB test görüntüsü kullanılmıştır. Buna ek olarak EndoScene’den\cite{endoscene} 60 görüntü, ETIS\cite{etis}’den 196, CVC-ColonDB’den\cite{colondb} ise 380 görüntü test için kullanılmıştır. Eğitim veri kümesinde aşırı öğrenmeyi engellemek için yatay çevirme ve rastgele döndürme veri çoğaltma teknikleri kullanılmıştır. Veri çoğaltma işlemi sonrası, eğitim veri kümesindeki görüntü sayısı 3870 olmuştur. Eğitim, doğrulama ve test veri kümelerindeki görüntüler 224x224 boyutuna sahip olacak şekilde yeniden boyutlandırılmıştır.
\subsection{Yöntem}
Derin öğrenme ile görüntü bölütleme modellerinin çoğunda beş katmanlı kodlayıcı kullanılmaktadır. Yüksek çözünürlüğe sahip ilk kodlayıcı katmanları girdi görüntüsü için daha düşük seviye bilgiler içermektedir. Düşük seviye olarak adlandırılan bu bilgiler görüntüdeki kenar bilgileri gibi çoğu görüntüde görülen ortak özelliklerdir. İleri kodlayıcı katmanları daha yüksek seviye bilgiler içermektedir. Bu özellikler bölütlenmesi amaçlanan nesne hakkında daha spesifik bilgiler içermektedir. Wu vd.\cite{pd} ilk kodlayıcı katmanların model başarısına katkısının daha az olmasının yanı sıra hesaplama maliyetlerinin de yüksek olduğunu belirtmişlerdir. Bunun yanında üçüncü katmandaki kodlayıcı katmanların birinci ve ikinci katmanın içerdiği düşük seviye bilgiye sahip olduğunu belirtmişlerdir. Bu bilgi doğrultusunda, modelimizde parametre sayısını artıran ancak bölütleme başarısına etkisi az olan ilk iki kodlayıcı katmanı tam ölçekli bağlantıda kullanmamayı seçerek parametre sayısını önemli ölçüde azalttık. Bu sayede parametre sayısı önemli ölçüde azaltılırken, bölütleme başarısı korunmuştur. Tıbbi görüntü içeren veri kümeleri genelde az veriye sahip olduğu için kodlayıcı katmanlarda ön eğitimli EfficientNetB0\cite{efficientnet} kullanılmaktadır. Bu mimarinin tercih edilmesinin sebebi daha az parametre ile yüksek başarı sonuçları vermesidir. Biz de çalışmamızda EfficientNetB0 kullandık. Bir çok derin öğrenme çalışması klasik evrişim blokları kullanmaktadır. Bunlar 3x3, 5x5, 7x7 gibi filtrelerden oluşmaktadır. Tıbbi görüntüler, özellikle kalın bağırsaktaki polipler, değişken şekil ve boyutlara sahip olduğu için klasik evrişim blokları kullanmak yerine asimetrik evrişim bloklarını\cite{asymmetric} kullanmayı tercih ettik. Aşağıda “Asimetrik evrişim bloğu (1)” formülü gösterilmiştir.
\begin{equation}
\label{denklem}
    evrisim(3x3)+evrisim(1x3)+evrisim(3x1)
\end{equation}
 
Bölütleme başarısını artırmak için özellik haritalarını ağırlıklandırarak önemli özellikleri ön plana çıkartan sıkma ve uyarma\cite{squeeze} bloğunu kullandık. Bu bloğun girdisi asimetrik bloğun çıktısıdır. U-Net\cite{unet} çalışması, kodlayıcı ve kod çözücü blokları arasındaki özellik haritaları bağlantılarını, aynı boyuttaki kodlayıcı ve kod çözücü katmanları arasında yapmaktadır. Çalışmamızda bu yaklaşım yerine tam ölçekli bağlantı\cite{unet3plus} yapısını kullandık. Tam ölçekli bağlantının amacı, düşük ve yüksek seviye bilgileri farklı ölçeklerde birleştirmektir. Bu sayede polip bölgelerindeki olası bilgi kaybı minimize edilmeye çalışılmıştır. Modelimizin parametre sayısını azaltmak için bir diğer yaklaşımımız her bir kodlayıcı katmanının çıktısını 64 filtreden geçirerek kod çözücü katmanlarında birleştirmektir. Bu sayede parametre sayısı önemli ölçüde azaltılmıştır. Modelimiz 40 devirde eğitilmiştir. Doğrulama kaybına bağlı olarak erken durdurma uygulanmıştır. Yığın sayısı 4 olarak belirlenmiştir. Optimizasyon algoritması olarak Adam kullanılmıştır. Öğrenme katsayısı 1e-4 olarak belirlenmiştir. Şekil \ref{model}’ de önerilen modelimizin genel yapısını görmekteyiz. Modelimiz toplamda 2.626.337 parametre gerektirmektedir.

\section{Sonuç ve Tartışma}
Çalışmamız U-Net\cite{unet}, U-Net++\cite{unet++}, SFA\cite{sfanet}, PraNet\cite{pranet} çalışmaları ile kıyaslanmıştır. Tablo \ref{parametre_sayısı}’ de karşılaştırılan modellerin ve bizim modelimizin toplam parametre sayısını görmekteyiz. Literatürde büyük kabul görmüş bu çalışmalara göre kullandığımız parametre sayısı oldukça azdır. Parametre sayısını kısmi kod çözücü yöntemi ile önemli ölçüde azalttık. Bunun yanında her bir kodlayıcı katmanın çıktısını kod çözücü katmanlarda birleştirmeden önce 64 filtreden geçirerek eğitilecek parametre sayısını azalttık. Dahası asimetrik evrişim blokları ile değişken boyutlar gösteren poliplerden daha iyi bilgiler çıkarttık. Sıkma ve uyarma bloğu ile önemli özellik haritalarını ön plana çıkartarak bölütleme başarısını artırdık. Tam ölçekli bağlantı yöntemi ile düşük ve yüksek seviye bilgileri farklı ölçeklerde bir araya getirerek havuzlama katmanında kaybolabilecek polip bilgilerini azaltmaya çalıştık. Şekil \ref{polyps}’de modelimizin örnek bölütleme sonuçlarına yer verilmiştir.   Tablo \ref{sonuclar}’de geliştirdiğimiz modelin beş farklı veri kümesindeki sonucunu görmekteyiz. Aynı zamanda diğer çalışmalarla olan kıyaslanması aynı tabloda yer almaktadır. Çalışmamız U-Net, U-Net++ ve SFA modellerinden tüm veri kümelerinde ve tüm metriklerde daha iyi sonuç vermiştir. Dice metriğine göre test kümesi olarak kullanılan ColonDB, EndoScene ve ETIS’de sırasıyla \%71.8, \%89.3 ve \%74.8 başarı elde edilmiştir. Bu sayısal değerlere göre çalışmamız tüm modellerden daha üstün sonuçlar elde etmiştir.

\bibliographystyle{IEEEtran}
\bibliography{ref.bib}

\begin{thebibliography}{10}
\providecommand{\url}[1]{#1}
\csname url@samestyle\endcsname
\providecommand{\newblock}{\relax}
\providecommand{\bibinfo}[2]{#2}
\providecommand{\BIBentrySTDinterwordspacing}{\spaceskip=0pt\relax}
\providecommand{\BIBentryALTinterwordstretchfactor}{4}
\providecommand{\BIBentryALTinterwordspacing}{\spaceskip=\fontdimen2\font plus
\BIBentryALTinterwordstretchfactor\fontdimen3\font minus
  \fontdimen4\font\relax}
\providecommand{\BIBforeignlanguage}[2]{{%
\expandafter\ifx\csname l@#1\endcsname\relax
\typeout{** WARNING: IEEEtran.bst: No hyphenation pattern has been}%
\typeout{** loaded for the language `#1'. Using the pattern for}%
\typeout{** the default language instead.}%
\else
\language=\csname l@#1\endcsname
\fi
#2}}
\providecommand{\BIBdecl}{\relax}
\BIBdecl

\bibitem{kvasir}
D.~Jha, P.~H. Smedsrud, M.~A. Riegler, P.~Halvorsen, T.~de~Lange, D.~Johansen,
  and H.~D. Johansen, ``Kvasir-seg: A segmented polyp dataset,'' in
  \emph{International Conference on Multimedia Modeling}.\hskip 1em plus 0.5em
  minus 0.4em\relax Springer, 2020, pp. 451--462.

\bibitem{unet}
O.~Ronneberger, P.~Fischer, and T.~Brox, ``U-net: Convolutional networks for
  biomedical image segmentation,'' in \emph{Medical Image Computing and
  Computer-Assisted Intervention -- MICCAI 2015}, N.~Navab, J.~Hornegger, W.~M.
  Wells, and A.~F. Frangi, Eds.\hskip 1em plus 0.5em minus 0.4em\relax Cham:
  Springer International Publishing, 2015, pp. 234--241.

\bibitem{resunet++}
D.~J. et~al., ``Resunet++: An advanced architecture for medical image
  segmentation,'' \emph{2019 IEEE International Symposium on Multimedia (ISM)},
  pp. 225--2255, 2019.

\bibitem{resnet}
K.~He, X.~Zhang, S.~Ren, and J.~Sun, ``Deep residual learning for image
  recognition,'' in \emph{2016 IEEE Conference on Computer Vision and Pattern
  Recognition (CVPR)}, 2016, pp. 770--778.

\bibitem{clinicdb}
J.~Bernal, F.~Śanchez, G.~Fernández-Esparrach, D.~Gil, C.~Rodríguez~de
  Miguel, and F.~Vilariño, ``Wm-dova maps for accurate polyp highlighting in
  colonoscopy: Validation vs. saliency maps from physicians,''
  \emph{Computerized Medical Imaging and Graphics}, vol.~43, 03 2015.

\bibitem{dunet}
D.~{Jha}, M.~A. {Riegler}, D.~{Johansen}, P.~{Halvorsen}, and H.~D. {Johansen},
  ``Doubleu-net: A deep convolutional neural network for medical image
  segmentation,'' in \emph{2020 IEEE 33rd International Symposium on
  Computer-Based Medical Systems (CBMS)}, 2020, pp. 558--564.

\bibitem{vgg}
K.~Simonyan and A.~Zisserman, ``Very deep convolutional networks for
  large-scale image recognition,'' \emph{CoRR}, vol. abs/1409.1556, 2015.

\bibitem{pranet}
D.-P. Fan, G.-P. Ji, T.~Zhou, G.~Chen, H.~Fu, J.~Shen, and L.~Shao, ``Pranet:
  Parallel reverse attention network for polyp segmentation,'' in \emph{Medical
  Image Computing and Computer Assisted Intervention -- MICCAI 2020}, A.~L.
  Martel, P.~Abolmaesumi, D.~Stoyanov, D.~Mateus, M.~A. Zuluaga, S.~K. Zhou,
  D.~Racoceanu, and L.~Joskowicz, Eds.\hskip 1em plus 0.5em minus 0.4em\relax
  Cham: Springer International Publishing, 2020, pp. 263--273.

\bibitem{colondb}
N.~Tajbakhsh, S.~R. Gurudu, and J.~Liang, ``Automated polyp detection in
  colonoscopy videos using shape and context information,'' \emph{IEEE
  Transactions on Medical Imaging}, vol.~35, no.~2, pp. 630--644, 2016.

\bibitem{endoscene}
D.~Vazquez, J.~Bernal, F.~J. Sanchez, G.~F. Esparrach, A.~M. Lopez, A.~Romero,
  M.~Drozdzal, and A.~C. Courville, ``A benchmark for endoluminal scene
  segmentation of colonoscopy images,'' \emph{CoRR}, vol. abs/1612.00799, 2016.

\bibitem{etis}
J.~Silva, A.~Histace, O.~Romain, X.~Dray, and B.~Granado, ``Toward embedded
  detection of polyps in wce images for early diagnosis of colorectal cancer,''
  \emph{International journal of computer assisted radiology and surgery},
  vol.~9, 09 2013.

\bibitem{unet++}
Z.~Zhou, M.~M.~R. Siddiquee, N.~Tajbakhsh, and J.~Liang, ``Unet++: Redesigning
  skip connections to exploit multiscale features in image segmentation,''
  \emph{IEEE Transactions on Medical Imaging}, vol.~39, no.~6, pp. 1856--1867,
  2020.

\bibitem{efficientnet}
M.~Tan and Q.~V. Le, ``Efficientnet: Rethinking model scaling for convolutional
  neural networks,'' in \emph{Proceedings of the 36th International Conference
  on Machine Learning}, vol.~97, 2019, pp. 6105--6114.

\bibitem{pd}
Z.~Wu, L.~Su, and Q.~Huang, ``Cascaded partial decoder for fast and accurate
  salient object detection,'' \emph{2019 IEEE/CVF Conference on Computer Vision
  and Pattern Recognition (CVPR)}, pp. 3902--3911, 2019.

\bibitem{squeeze}
J.~Hu, L.~Shen, and G.~Sun, ``Squeeze-and-excitation networks,'' in \emph{2018
  IEEE/CVF Conference on Computer Vision and Pattern Recognition}, 2018, pp.
  7132--7141.

\bibitem{asymmetric}
X.~Ding, Y.~Guo, G.~Ding, and J.~Han, ``Acnet: Strengthening the kernel
  skeletons for powerful cnn via asymmetric convolution blocks,'' in \emph{The
  IEEE International Conference on Computer Vision (ICCV)}, October 2019.

\bibitem{unet3plus}
H.~Huang, L.~Lin, R.~Tong, H.~Hu, Q.~Zhang, Y.~Iwamoto, X.~Han, Y.-W. Chen, and
  J.~Wu, ``Unet 3+: A full-scale connected unet for medical image
  segmentation,'' \emph{ICASSP 2020 - 2020 IEEE International Conference on
  Acoustics, Speech and Signal Processing (ICASSP)}, pp. 1055--1059, 2020.

\bibitem{sfanet}
Y.~Fang, C.~Chen, Y.~Yuan, and K.-y. Tong, ``Selective feature aggregation
  network with area-boundary constraints for polyp segmentation,'' in
  \emph{Medical Image Computing and Computer Assisted Intervention -- MICCAI
  2019}.\hskip 1em plus 0.5em minus 0.4em\relax Cham: Springer International
  Publishing, 2019, pp. 302--310.

\end{thebibliography}

\end{document}